\documentclass[prb,superscriptaddress,twocolumn,floatfix]{revtex4}
\usepackage{amsmath, amsthm, amssymb}
\usepackage{bm} 
\usepackage{amsfonts} 
\usepackage{graphicx} 
\begin{document}

\title{The prediction of future from the past: an old problem
from a modern perspective}

\author{F. Cecconi}\affiliation{Istituto dei Sistemi Complessi,
  Consiglio Nazionale delle Ricerche, Via dei Taurini 19, I-00185
  Rome, Italy}

\author{M. Cencini}\affiliation{Istituto dei Sistemi Complessi,
  Consiglio Nazionale delle Ricerche, Via dei Taurini 19, I-00185
  Rome, Italy}

\author{M. Falcioni}\email{massimo.falcioni@roma1.infn.it}
\affiliation{Dipartimento di Fisica, Universit\`a ``Sapienza'',
  Piazzale A. Moro 2, 00185 Rome Italy}

\author{A. Vulpiani}\affiliation{Dipartimento di Fisica, Universit\`a
  ``Sapienza'', Piazzale A. Moro 2, 00185 Rome Italy}\affiliation{Istituto dei Sistemi Complessi, Consiglio
  Nazionale delle Ricerche, Via dei Taurini 19, I-00185 Rome, Italy}

\begin{abstract}
The idea of predicting the future from the knowledge of the past is
quite natural when dealing with systems whose equations of motion are
not known.  Such a long-standing issue is revisited in the light of
modern ergodic theory of dynamical systems and becomes particularly
interesting from a pedagogical perspective due to its close link with
Poincar\'e's recurrence.  Using such a connection, a very general
result of ergodic theory - Kac's lemma - can be used to establish the
intrinsic limitations to the possibility of predicting the future from
the past. In spite of a naive expectation, predictability results to
be hindered rather by the effective number of degrees of freedom of a
system than by the presence of chaos.  If the effective number of
degrees of freedom becomes large enough, regardless the regular or
chaotic nature of the system, predictions turn out to be practically
impossible.  The discussion of these issues is illustrated with the
help of the numerical study of simple models.
\end{abstract}

\maketitle

\section{Introduction}

Predicting the future state of a system had always been a natural
motivation for science development, with applications such as weather
forecasting and tidal prediction.  Understanding the limitations to
the predictability of a system evolution is therefore crucial.

In deterministic systems, where the future is uniquely determined by
the present, two main approaches to the predictability problem can be
addressed.  The first refers to systems whose evolution laws are
known, either in terms of differential or difference equations. In
this case, the predictability is mainly limited by the presence of
sensitivity to initial conditions (deterministic chaos), which as
taught in dynamical system courses, is characterized by the Lyapunov
exponent.  The second approach refers to phenomena whose governing
laws are not known, but whose evolution can be measured and recorded.
In such a case, the best practical strategy is to use the past, as a
full-scale model of the system, to make predictions on the future
evolution.

The present paper discusses at an introductory level the latter
method, which was developed in the framework of nonlinear time series
analysis.\cite{time-series,Abarbanel,Kantz1997} This topic is seldom
included in basic courses and is closely related to an apparently
distant classical theme --- the Poincar\'e recurrences.\cite{Poincare}
Surprisingly, although simple to establish,\cite{Nicolis1998} such a
connection has been overlooked even by specialists, as recently
remarked by Altmann and Kantz.\cite{Altmann2005} Such a link also
allows us to clarify the practical role of theoretical concepts as the
attractor dimension of a dynamical system. Indeed, as we shall see,
when the evolution laws are unknown, the actual constraints to our
prediction capabilities are rather set by the number of degrees of
freedom (attractor dimension) than by the presence of chaos.  This
fact is often overlooked in favor of the widespread folklore of the
\textit{butterfly effect}.\cite{Gleick} In this respect, it is
important to stress that such limitations to predictability are a
consequence of rather general results of ergodic and dynamical-system
theory. Although the main ideas had been already put forward by
Boltzmann,\cite{Cercignani} many misguided applications of nonlinear
time series analysis appeared in the literature after the rediscovery
of chaos (see, e.g., Ref.~\onlinecite{Ruelle1989}).

Likely, one of the main reasons for excluding this topic from basic
courses is the necessity to introduce advanced technical
tools\cite{Kantz1997} as, for example, the embedding
technique.\cite{Takens1981,Kantz1997} Therefore, here, we present the
problem in its simplest formulation. Often, when recording the
evolution of a system with unknown dynamics, not all the variables
necessary to identify the states or even their number are
known. Moreover, if luckily we know them, we can access only one or a
few scalar functions of them, typically affected by measurement
errors. Throughout this paper, we will disregard all these technical
difficulties (which can be to a large extent handled with specific
techniques~\cite{Kantz1997}) and assume that the necessary variables
can be recorded with arbitrary precision. Even with such ideal working
hypothesis, the above mentioned fundamental constraints to
predictability are unavoidable.
 
The material is organized as follows. In Sect.~\ref{sec:2}, after some
historical notes, we introduce the method of analogues as the simplest
procedure to predict the future from past time
series. Sect.~\ref{sec:3} introduces the model system used to clarify
the main issues.  In Sect.~\ref{sec:4}, we discuss the link between
analogues and Poincar\'e recurrences, and show how the actual
limitations to predictability from data stems from the effective
number of degrees of freedom. Sect.~\ref{sec:5} discusses two cases
where the method works successfully, one is illustrated by a numerical
example and the other refers to the important practical problem of
tidal predictions. Finally, Sect.~\ref{sec:6} is devoted to
conclusions.

\section{The method of analogues\label{sec:2}}

``If a system behaves in a certain way, it will do again'' seems a
rather natural claim when referred, for instance, to the diurnal and
seasonal cycles, also supported by biblical tradition: \textit{What
  has been will be again, what has been done will be done again; there
  is nothing new under the sun} [the Qohelet's Book 1:9 NIV]. This
idea, together with the belief in determinism ({\it from the same
  antecedents follow the same consequents}), is at the basis of
prediction methods.  However, as Maxwell argued:\cite{Maxwell}
\textit{It is a metaphysical doctrine that from the same antecedents
  follow the same consequents.$[\ldots]$ But it is not of much use in
  a world like this, in which the same antecedents never again concur,
  and nothing ever happens twice.$[\ldots]$ The physical axiom which
  has a somewhat similar aspect is ``That from like antecedents follow
  like consequents.''}  These words no more surprise the scientists,
aware, by now, of the almost exceptional character of periodic
behaviors and of the ubiquitous presence of irregular evolutions due
to deterministic chaos; but at that time they constituted a rupture
with the tradition.

In spite of Maxwell authoritative opinion, until World War I, weather
forecasters substantially used empirical implementations of the naive
idea, exploiting their experience and memory of past similar
``patterns'' (roughly surfaces of discontinuity between warm and cold
air masses) to produce weather map predictions.\cite{Lynch2006} In the
preface to his seminal book {\it Weather Prediction by Numerical
  Process}, Richardson criticizes the empirical approaches and,
through an argument similar to that by Maxwell,\cite{Richardson1922}
contends that for weather forecasting it is much more useful
integrating the partial differential (namely the
thermo-hydrodynamical) equations ruling the atmosphere.  Although, as
history witnessed, the successful approach to predictions is that
foreseen by Richardson, it is interesting to discuss the range of
applicability of predictions based on the past evolution of a
deterministic system.

A mathematical formulation of the idea was due to Lorenz and it is
called \textit{method of analogues},\cite{Lorenz1969a,Lorenz1969b}
which can be considered as the most straightforward approach to
predictability in the absence of a detailed knowledge of the physical
laws.

\begin{figure}[hbt]
\includegraphics[clip=true,width=0.43\textwidth]{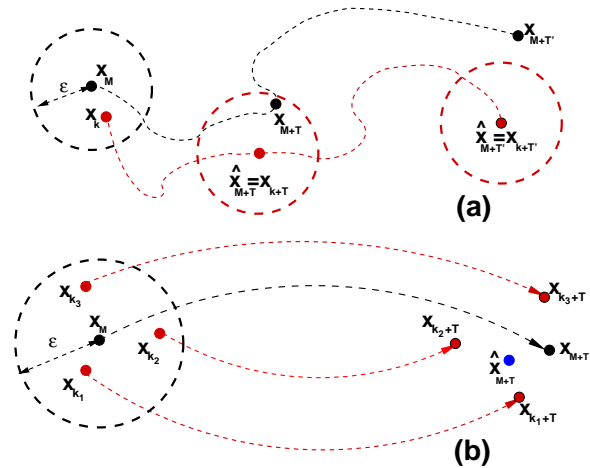}
\caption{(Color online) Sketch of the method of analogues: (a)
  illustration of Eq.~(\ref{eq:simplepred}) and of the error growth;
  (b) generalization of the method to more than one analogue. In
  particular, if $N_a$ analogues, $\{\bm x_{k_n}\}_{n=1}^{N_a}$ are
  found (\ref{eq:simplepred}) can be replaced by $ \hat{\bm x}_{M+T}
  =\sum_{n=1}^{N_a} \mathbb{E}_n {\bm x}_{k_n +T}$ where the matrices
  $\mathbb{E}_n$ can be computed by suitable interpolations.
\label{fig:analogue}}
\end{figure}

In its simplest description, the method works as follows. Assume that
the known state $\bm x(t)$ of a process can be sampled at times
$t_k=k\Delta t$ with arbitrary precision. The sampling interval
$\Delta t$ is also assumed to be arbitrary but not too short. We
collect the sequence of states $\bm x_k=\bm x(t_k)$ with $k=1,\ldots
M$. If from the present state $\bm x_M$, we would like to forecast the
future $\bm x_{M+T}$ at time $t_{M+T}$ ($T\geq 1$), the basic idea is
to search in the past $(\bm x_1,\bm x_2,\ldots,\bm x_{M-1})$ that
state, say $\bm x_k$, most similar to $\bm x_M$, and to use its
consequents as proxies for the future evolution of $\bm
x_M$. Mathematically, we require that $|\bm x_k - \bm x_M|\leq
\epsilon$, and we dub $\bm x_k$ a $\epsilon$-\textit{analogue} to $\bm
x_M$. If the analogue were perfect ($\epsilon=0$) the system (being
deterministic) would be surely periodic and the prediction trivial:
$\bm x_{M+T}\equiv \bm x_{k+T}$ for any $T$.  If it were not perfect
($\epsilon >0$), we could use the forecasting recipe
\begin{equation}
\hat{\bm x}_{M+T}= \bm x_{k+T}\,,
\label{eq:simplepred}
\end{equation}
as \textit{from like antecedents follow like consequents} (see
Fig.~\ref{fig:analogue}a).  For the prediction (\ref{eq:simplepred})
to be meaningful, the analogue $\bm x_k$ must not be a near-in-time
antecedent.  When more than one analogue can be found, the
generalization of (\ref{eq:simplepred}) is obvious, see
Fig.~\ref{fig:analogue}b.

Once a ``good'' analogue (meaning $\epsilon$ reasonably small) has
been found, the next step is to determine the accuracy of the
prediction (\ref{eq:simplepred}), namely the difference between
the forecast and the actual state $|\hat{\bm x}_{M+T}-\bm x_{M+T}|$.  In
practice, the $\epsilon$-analogue is the present state with an uncertainty,
$\bm x_k=\bm x_M+\delta_0$ ($\delta_0\leq \epsilon$), and the
prediction (\ref{eq:simplepred}) can be considered acceptable until 
the error $\delta_{T}=|\bm x_{M+T}-\bm x_{k+T}|$ remains
below a tolerance $\Delta$, dictated by the practical needs. The
predictability time $\widehat{T}=\widehat{T}(\delta_0,\Delta)$ is then defined
by requiring $\delta_{T}\lesssim \Delta$ for $T\leq \widehat{T}$.

Accuracy and predictability time $\widehat{T}$ are clearly related to
(possible) sensitivity to initial conditions, as pioneered by Lorenz
himself.\cite{Lorenz1963} As taught in basic dynamical system courses,
chaotic evolutions exponentially amplify an infinitesimal error:
\begin{equation}
\label{eq:lyap}
\delta_T \simeq
\delta_0 e^{\lambda_1 T}\,,
\end{equation}
 $\lambda_1$ being the maximal Lyapunov exponent.\cite{libro} For a
gentle introduction to Lyapunov exponents the reader may refer to
Ref.~\onlinecite{desouza1990}.  Therefore, given a good analogue the
prediction will be $\Delta$-accurate up to a time
\begin{equation}
\widehat{T}(\delta_0,\Delta)\approx \frac{1}{\lambda_1}
\ln\frac{\Delta}{\delta_0}\,.
\end{equation}
Strictly speaking, for the above equation to be valid, both $\delta_0$
and $\Delta$ must be very small.\cite{libro}
It is worth remarking that the evaluation of the error growth rate
(\ref{eq:lyap}) provides, at least in principle, a way to determine
the Lyapunov exponent $\lambda_1$ from a long time series.

Conversely, deterministic non-chaotic systems are less sensitive to
initial conditions: the error grows polynomially in time, and usually,
$\widehat{T}(\delta_0,\Delta)$ results to be longer than that of
chaotic systems, making long term predictions possible.

For those familiar with chaotic systems, we apparently reached the
obvious conclusion that the main limit to predictions based on
analogues is the sensitivity to initial conditions, typical of chaos.
But, as realized by Lorenz himself, the main issue is to find good
(small $\epsilon$) analogues:\cite{Lorenz1969b} \textit{In practice
  this procedure may be expected to fail, because of the high
  probability that no truly good analogues will be found within the
  recorded history of the atmosphere.}  He also pointed out that the
very drawback of the method stems from the need of a very large data
set,\cite{Lorenz1969a} independently of the presence of chaos.

It is worth concluding this historical presentation with a brief
comment on the application of the method of analogues in the original
Lorenz's work.\cite{Lorenz1969a} Lorenz was strongly supporting
weather forecasting based on solving the (approximate) equations of
the atmosphere, as outlined by Richardson. He realized that the
intrinsic limits to weather forecasting cannot be established by
estimating the intrinsic error growth of these solutions.  This work
represents the first attempt to estimate the Lyapunov exponent from
data, pioneering the modern time series analysis.\cite{Wolf}
Unfortunately, he also realized that, the true Lyapunov exponent of
the atmosphere cannot be estimated from data, as good analogues cannot
be found and the \textit{difference between mediocre analogues may be
  expected to amplify more slowly than the difference between good
  analogues, since the non linear effects play a greater role when the
  errors are large}.

\section{Study of a simple model\label{sec:3}}

The difficulties in finding good analogues can be
quantified by studying analogue statistics. As an
illustrative example, we compute numerically the probability of
finding $\epsilon$-analogues to a state in a simple model system
introduced by Lorenz in 1996,\cite{Lorenz1996} hence called Lorenz-96
model.  It consists of the following nonlinearly coupled ordinary
differential equations
\begin{equation}
\label{l96}
\frac{ \textrm{d} X_n} {\textrm{d} t} \!=\! 
X_{n-1}(X_{n+1}\!-\!X_{n-2}) -X_{n}+ F\, ,
\end{equation}
where $n\!=\!1,\ldots, N$ and periodic boundary conditions ($X_{N\pm
  n}=X_{\pm n}$) are assumed.  The variables $X_n$ may be thought of
as the values of some atmospheric representative observable along the
latitude circle, so that Eq.~(\ref{l96}) can be regarded as a
one-dimensional caricature of atmospheric motion.\cite{Lorenz1996} The
quadratic coupling conserves energy, $\sum_n X^2_n$. In the presence
of forcing $F$ and damping $-X_n$, the energy is only statistically
conserved. The motion is thus confined to a bounded region of
$\mathbb{R}^N$. Moreover, dissipation constraints the trajectories to
evolve onto a subset of this region possibly with dimension $<N$,
namely an attractor (fixed points, limit cycles or a \textit{strange
  attractor} if the dynamics is chaotic). The dynamical features are
completely determined by the forcing strength $F$ and by the system
dimensionality $N$. In particular, for $F > 8/9$ and $N \ge 4$ the
system displays chaos with exponential separation of nearby initial
conditions.\cite{lore2scale}

In principle, the statistics of the analogues of system (\ref{l96})
can be determined according to the following procedure.  Given a state
of the system $\bm x_{M}$ on the attractor, we have to consider its
precursors $(\bm x_1,\ldots,\bm x_{M-1})$ along the trajectory ending
in $\bm x_M$ sampled at regular time intervals of duration $\Delta t$,
$\bm x_i=\bm x(t_i=i \Delta t)$. Hence, the $\epsilon$-analogues of
$\bm x_M$ are those states $\bm x_j$ such that $\vert \bm x_j- \bm
x_M\vert \leq \epsilon$. Finally, the fraction of
$\epsilon$-analogues,
\begin{equation}
\label{anafrac}
C_M(\epsilon)= \frac{1}{M-1} \sum_{j=1}^{M-1} 
\Theta( \epsilon -\vert\bm x_j- \bm x_M\vert)  \, ,
\end{equation}
provides an estimate of the probability to find $\epsilon$-analogues
to $\bm x_M$ as a function of both the desired degree of similarity
$\epsilon$ and the length of the history $M$ we recorded.  Being
interested in typical behaviors and not just in the properties around
a specific state $\bm x_M$, it is convenient to average
$C_M(\epsilon)$ over $r$ independent reference states.  Therefore,
instead of considering only the end point $\bm x_M$, we select $r$
states $\{\bm x_\star^{k}\}^r_{k=1}$ along the trajectory, well spaced
in time to be considered independent configurations on the attractor,
and we replace (\ref{anafrac}) by the average fraction of
$\epsilon$-analogues
\begin{equation}
\label{grasspro}
C_{r,M}(\epsilon)= \frac{1} {Mr} \sum_{k=1}^{r}\sum_{j=1}^M 
\Theta( \epsilon -\vert{\bm x}_j- {\bm x}_\star^{(k)}\vert)  \, .
\end{equation} 
As in our case we know the evolution laws (\ref{l96}), it is not
really necessary to look at the backward time series of the reference
states. In practice, we can select the $\{\bm x_\star^{k}\}^r_{k=1}$
and look at their forward $\epsilon$-analogues.

The latter procedure is used to produce Fig.~\ref{corrX}, where we
show $C_{r,M}(\epsilon)$ obtained with ${r}=10^{3}$ reference states
and different lengths $M$ of the time series, from $10^3$ to $10^7$.
Of course, when the degree of similarity $\epsilon$ becomes larger
than the attractor size, say $\epsilon_{\mathrm{max}}$, the fraction
$C_{r,M}(\epsilon)$ saturates to $1$. Therefore, it is meaningful to
normalize the degree of similarity by $\epsilon_{\textrm{max}}$.  As
for the dynamics (\ref{l96}), the forcing is fixed to $F=5$ and we
consider two system sizes $N=20$ and $N=21$. In both cases the system
is chaotic.  While for $N=21$ analogues can be found with reasonable
probability even for small values of $\epsilon\,(\lesssim 10^{-4}
\epsilon_{\mathrm{max}})$, for $N=20$ analogues are found only for
large values of $\epsilon\,(\gtrsim 10^{-2} \epsilon_{\mathrm{max}})$,
even for $M=10^{7}$. The solid lines in Fig.~\ref{corrX} indicate that
for $\epsilon \ll \epsilon_{\textrm{max}}$ the probability to find an
analogue is fairly well approximated by a power law
\begin{equation}
C_{r,M} (\epsilon) \propto \epsilon^{D_A}\,.
\label{eq:plaw}
\end{equation}
In particular, we find $D_A\simeq 3.1$ and $D_A\simeq 6.6$ for $N=21$
and $N=20$, respectively. Therefore, the exponent $D_A$ quantifies
the difference between the two cases: upon lowering $\epsilon$, the
probability to find $\epsilon$-analogues with $N=20$ becomes about
$\epsilon^{3.5}$ times smaller than with $N=21$.

The probability to find $\epsilon$-analogues is expected to decrease
upon increasing the number of degrees of freedom $N$, as more
constraints on the single components of the state vector should be
satisfied. In this perspective, the above result seems at odds with
intuition unless the exponent $D_A$ in (\ref{eq:plaw}) is interpreted
as the ``effective'' number of degrees of freedom.

We end this section by warning the reader that the counter-intuitive
inequality $D_A(N=21)<D_A(N=20)$ is a peculiar consequence of the
choice of the parameters $F$ and $N$.\cite{lore2scale} Generally,
$D_A$ is expected to increase with $N$.\cite{libro} Here, we made this
choice to emphasize the importance of the effective number of degrees
of freedom that, in general, is not trivially related to (and can be
much smaller than) the number of variables $N$. As we shall see in the
next section, $D_A$ is nothing but the attractor dimension, a measure
of the effective number of degrees of freedom.

\section{Degrees of freedom, recurrence times and analogues
\label{difficult}\label{sec:4}}

In this Section we recall some basic notions of ergodic dynamical systems and
underline their connections with the analogues. In
particular, we link the difficulty of finding analogues to the
presence of long recurrence times.

\subsection{The role of dimensions}

The founding principle of ergodic theory is that the long-time
statistical properties of a system can be equivalently described in
terms of the invariant (time-independent) probability, $\mu$, such
that $\mu(\sigma)$ is the probability of finding the system in any
specified region $\sigma$ of its phase space.  The phase space of a
system described by $N$ degrees of freedom is a region of
$\mathbb{R}^N$, that is a $N$-dimensional space.

If the evolution conserves phase-space volumes (as in the Hamiltonian
motion of classical systems) then the probability $d\mu({\bm x})$ of
finding the state in a small region of volume $dV\!\!,$ as defined in
elementary geometry, around ${\bm x}$ is proportional to $d V$,
i.e. to the Lebesgue measure of that region. In dissipative systems,
phase-space volumes are contracted on average and the invariant
probability $d\mu({\bm x})$ is not proportional to $d V$, but
concentrates on a set (the attractor) $A\subset \mathbb{R}^N$ of
dimension $D_A < N$. Slightly more formally, the dimension $D_A$ describes the 
small scale ($\ell \ll 1$) behavior of  the
probability $\mu \bigl(B^N_{\bm y}(\ell) \bigr)$ of finding points
$\bm x\in A$ which are in the $N$-dimensional sphere of radius $\ell$
around $\bm y$:
\begin{equation}
\label{frac-dim}
\mu \bigl ( B^N_{\bm y}(\ell) \bigr ) = \int_{B^N_{\bf y}(\ell)} d\mu({\bm
  x}) \sim \ell^{D_A} \,.
\end{equation}
Therefore, the trajectories of dissipative systems are effectively
described by a number $D_A<N$ of degrees of freedom, though defined in
a $N$-dimensional space.

For a non-integer $D_A$ attractor and probability are said to be
fractal. In general, attractors are non-homogeneous with $D_A$, in
Eq.~(\ref{frac-dim}), depending on $\bm y$, and an infinite set of
dimensions is needed to fully characterize the invariant probability
--- we speak of multifractal objects.\cite{Paladin} For the sake of
our discussion, these technical complications can be ignored, and the
attractor can be assumed homogeneous and characterized by a single
dimension $D_A$.

Upon reconsidering $C_M(\epsilon)$ defined in Eq.~(\ref{anafrac}), we
see that it is nothing but the fraction of time the trajectory spends
in a sphere or radius $\epsilon$ centered in ${\bm x}_M$. For large
$M$, as a consequence of ergodicity, $C_M(\epsilon)$ gives the
probability of finding the system in that sphere, and the quantity
(\ref{grasspro}) is an averaged probability.  Therefore, for
sufficiently large $M$ and small $\epsilon$, Eq.~(\ref{frac-dim})
implies
\begin{equation}
\label{c-df}
C_{r, M} (\epsilon) \approx \langle \mu(\epsilon) \rangle 
 \sim \epsilon^{D_A} \,\,.
\end{equation}
Strictly speaking, in Eq.~(\ref{c-df}) the right exponent should be
the correlation dimension $D_2$, which controls the small scale
asymptotics of the probability to find two points on the attractor at
distance $\leq \epsilon$.\cite{Paladin,libro} Thanks to the homogeneity
assumption, however, we have $D_2 \simeq D_A$.

\begin{figure}[hbt]
\includegraphics[clip=true,width=0.44\textwidth]{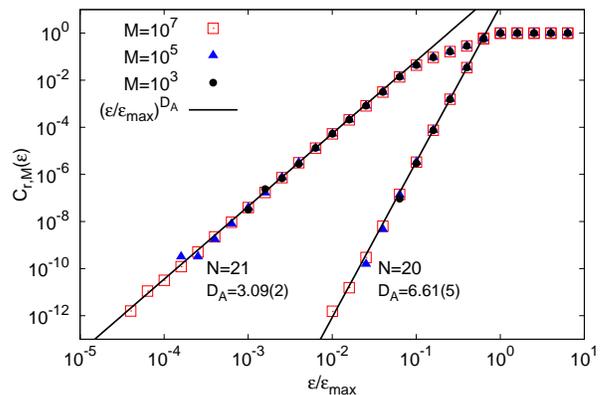}
\caption{(Color online) $C_{{r},M}(\epsilon)$
  vs. $\epsilon/\epsilon_{\textrm{max}}$ for $F=5$, $N=20$ and $N=21$;
  the reference states are ${r}=1000$ and different values of $M$
  ranging from $10^3$ to $10^7$ are considered. The solid lines are
  the fits of the data by means of
  relation~(\ref{c-df}).\label{corrX}}
\end{figure} 

Relation (\ref{c-df}) links the observed behavior (\ref{eq:plaw}) in
Fig.~\ref{corrX} to the attractor dimension, showing that the limiting
factor to find good analogues is the attractor dimension, which
quantifies the number of ``active'' degrees of freedom of the system.
For those accustomed to chaotic systems, this result is rather
obvious as $C_{r,M}(\epsilon)$ in (\ref{grasspro}) provides a standard
approximation\cite{Kantz1997} to the correlation sum, $2/(M(M-1))
\sum_{i,j>i} \Theta(\epsilon-|\bm x_i -\bm x_j|)$, at the basis of the
Grassberger and Procaccia method to determine the correlation
dimension $D_2$.\cite{Grassberger1983} Indeed, the correlation sum is an
unbiased estimator of the probability $P_2(\epsilon)$ to find two
randomly chosen points on the attractor (using a long trajectory on
it) at a distance $\leq \epsilon$. For small $\epsilon$,
$P_2(\epsilon)\sim \epsilon^{D_2}$ and thus the correlation dimension
can be estimated.\cite{libro}

\subsection{Poincar\'e recurrence theorem and Kac's lemma}
The quantity $C_M(\epsilon)$, besides approaching (for large $M$) the
probability to find the system state $\epsilon$-close to $\bm x_M$,
relates to the average time interval $\overline{ \tau}_{_R}$ between
two consecutive $\epsilon$-analogues of $\bm x_M$, which is given by
\begin{equation}
\label{tau}
\overline{ \tau}_{_R}= \frac{(M-1)\Delta t}{\mathcal{M}(\epsilon)}\,,
\end{equation}
$\mathcal{M}(\epsilon)$ being the number of $\epsilon$-analogues in
the interval $[t_1:t_{M-1}]$. As by definition
$C_M(\epsilon)=\mathcal{M}(\epsilon)/(M-1)$ we have
\begin{equation}
\label{c-tau}
C_M(\epsilon) = {\Delta t}/ \overline {\tau}_{_R}  \,\, ,
\end{equation}
actually this is a classical result of the ergodic theory --- known as
Kac's lemma.

To clarify this connection, it is worth recalling the Poincar\'e
recurrence theorem\cite{Poincare} stating that, in Hamiltonian
systems with a bounded phase space $\Omega$, the trajectories exiting
from a generic set $\sigma \subset \Omega$ will return back to
$\sigma$ infinitely many times. The theorem holds for almost all
points in $\sigma$ except for a possible subset of zero probability.
In general, it applies to the class of systems with volume-preserving
dynamics in phase-space, of which Hamiltonian ones are a particular
sub-class. Actually, although often not stressed in elementary
courses, it can be straightforwardly extended to dissipative ergodic
systems provided one only considers initial conditions on the
attractor and ``zero probability'' is interpreted with respect to the
invariant probability on the attractor.

Poincar\'e theorem merely proves that a trajectory surely return to
the neighborhood of its starting point, but does not provide
information about the time between two consecutive recurrences --- the
Poincar\'e recurrence time. The latter is crucial to the method of
analogues because long recurrence times critically spoil its
applicability (see Eq.~(\ref{c-tau})).

 To estimate the average recurrence time, let us assume that an
 infinitely long sequence of states can be stored.  Without loss of
 generality, we consider a discrete time sequence $\bm x_k=\bm
 x(k\Delta t)$ ($k=0,\ldots,\infty$) of states generated by a
 deterministic evolution from the initial condition ${\bm x}_0$.
 Given a set $\sigma$ including ${\bm x}_0$, the recurrence time of
 $\bm x_0$ relative to $\sigma$, $\tau_{\sigma}({\bm x}_0)$, can be
 defined as the minimum $k$ such that $\bm x_k$ is again in $\sigma$
\begin{equation}
\tau_{\sigma}({\bm x}_0)= \inf_{k}\{k\ge 1 |{\bm x}_0 
\in \sigma \; \mathrm{and}\; {\bm x}_k \in \sigma \} \;,
\end{equation} 
note that we are using dimensionless times with $\Delta
  t =1$.  The mean recurrence time relative to $\sigma$, $\langle
\tau_{\sigma} \rangle$, can then be computed as
\begin{equation}
\label{ave-tau}
\langle \tau_{\sigma} \rangle 
=\frac{1}{{\mu(\sigma)}} \int_{\sigma} d \mu({\bm x}) 
\tau_{\sigma}({\bm x})\,, 
\end{equation}
$\mu$ being the invariant probability with respect to the dynamics,
defined in the previous subsection.  For ergodic dynamics, a classical
result known as Kac's Lemma states that:\cite{Kac1959}
\begin{equation}
\label{kac1}
\int_{\sigma} d \mu({\bm x}) \tau_{\sigma}({\bm x}) =1
\quad \textrm{so that} \quad
\langle \tau_{\sigma} \rangle= {1}/ {{\mu(\sigma)}} \,,
\end{equation}
namely the average recurrence time to a region $\sigma$ is just the
inverse of the probability of that region. We stress that (\ref{kac1})
is a straightforward consequence of ergodicity.\footnote{We sketch the
  derivation in the simple case of a discrete-state system as obtained
  partitioning the phase-space of given system into a discrete number
  of regions $\lbrace \rho _j \rbrace$, each one defining a
  coarse-grained state $j$, such that the probability $p_j$ of a state
  $j$ is given by $p_j = \mu(\rho_j)$.  We can focus on a state
  $s$. Assume that during the time interval $\mathcal{T}$, a system
  trajectory has visited $\mathcal{N}_s$ times the $s$-th state
  (representing here the set $\sigma$). Denoting with $\tau_s(i)$ the
  time taken by the $i$-th return to the state $s$, the mean
  recurrence time is $\sum_{i=1}^{\mathcal{N}_s}
  \tau_s(i)/\mathcal{N}_s = \mathcal{T}/\mathcal{N}_s$, that is the
  inverse of the fraction of time spent by the system in the $s$-th
  state, $f_s = \mathcal{N}_s/\mathcal{T}$.  Invoking ergodicity, in
  the limit $\mathcal{T} \to \infty$ we have
  $\mathcal{T}/\mathcal{N}_s \to \langle \tau_s\rangle $ and $f_s\to
  p_s$ where $p_s$ is the probability of the $s$-th state (i.e. the
  probability $\mu(\sigma)$) so that $\langle \tau_s\rangle= 1/p_s$,
  which represents (\ref{kac1}).  Substituting $\mathcal{T}$ and
  $\mathcal{N}_s$ with $M-1$ and $\mathcal{M} (\epsilon)$,
  respectively, we obtain Eqs.~(\ref{tau}) and (\ref{c-tau}), with
  $\Delta t = 1$, so that for large $M$, Eq.~(\ref{c-tau}) gives again
  Kac's result. }

In a system with phase-space volume preservation (those for which the
Poincar\'e theorem is typically invoked) with $N$ degrees of
freedom, if $\sigma$ is an hypercube of linear size $\epsilon$ one has
\begin{equation}
\label{kac2}
\mu(\sigma) \sim \Bigl(\frac{\epsilon}{L} \Bigr) ^{N}
\quad \textrm{and} \quad
\langle \tau_{\sigma} \rangle \sim \Bigl(\frac{L}{\epsilon} \Bigr) ^{N} \,\, ,
\end{equation}
where $L$ is the typical excursion of each component of ${\bm x}$.
Thus the mean return time exponentially grows with $N$.  Consequently
in a macroscopic body ($N \gg 1$), $ \langle \tau_{\sigma} \rangle $
is astronomically large, for any $\sigma$.  The result (\ref{kac2}) is
surely positive for the validity of statistical mechanics, as
recognized by Boltzmann himself who (without knowing Kac's Lemma) replied to
Zermelo criticism to irreversibility: \textit{Of course if one waits
  long enough, the initial state will eventually recur, but the
  recurrence time is so long that there is no possibility of ever
  observing it}.\cite{Cercignani} But it is dramatically negative for
the possibility to find analogues in high dimensional systems.

In the case of ergodic dissipative systems, where the coarse-grained
probabilities are ruled by the dimension $D_A$ (Cfr. Eq.~(\ref{frac-dim})), 
Kac's result (\ref{kac2}) applies with $N$ replaced by $D_A$.

We conclude this digression on Poincar\'e recurrences by noting that
the limitations to find the analogues set by relation~(\ref{kac2}) is
unrelated to chaos.  For instance, Eq.~(\ref{kac2}) also applies to a
chain of $n$ harmonic oscillators with incommensurable frequencies,
a system with regular (quasiperiodic) behavior. Strictly
speaking, such a system is not ergodic in the whole angle-action phase
space, but in the space of angles only. Therefore in Eq.~(\ref{kac2})
instead of $N=2n$ one has to set $N=n$. \cite{Hemmer1958, Kac-remark}

\subsection{Consequence of Kac's Lemma}
 
The above results allow us to quantify Lorenz's pessimism with respect
to the number of data necessary for finding good analogues in the
atmosphere.\cite{Lorenz1969a} Clearly, we must require $M \Delta t
\gtrsim \overline {\tau}_{_R}$, which from (\ref{c-tau}) implies $M
\gtrsim 1/C_M(\epsilon)$. Then using (\ref{c-df}), we can realize that
the minimum length of the time series is
\begin{equation}
\label{M-df}
M \sim \Bigl(\frac{L}{\epsilon} \Bigr)^{D_A} \,\, ,
\end{equation}
$L$ being the typical excursion of each component of 
${\bm  x}$.

\begin{figure}[hbt]
\includegraphics[clip=true,width=0.44\textwidth]{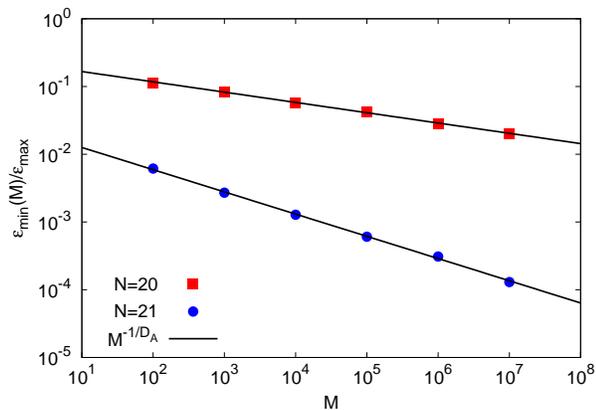}
\caption{(Color online) $\epsilon_{min}/\epsilon_{max}$ vs. $M$. The
  parameters of the model are the same as in Fig.~\ref{corrX}: $F=5$,
  $N=20$ and $N=21$; the reference states are ${r}=1000$. The solid
  lines are the fits of the data by means of
  relation~(\ref{M-df}).\label{epsM} }
\end{figure}

Equation~(\ref{M-df}) implies that, at least in principle, the method
can work for deterministic systems having an attractor of finite
dimension provided the time series is suitably long. However, the
exponential dependence on $D_A$ in Eq.~(\ref{M-df}) imposes, upon
putting the numbers, too severe constraints even if we content
ourselves of a poor precision, i.e. not too small
$\epsilon$-analogues. For instance in Fig.~\ref{epsM}, we show how the
distance between a reference point and its best analogue
($\epsilon_{\mathrm{min}}$) scales with $M$.  We see that for
$\epsilon_{\mathrm{min}}/\epsilon_{\mathrm{max}}=10^{-2}$ a sequence
of $10^2$ points is sufficiently long in the case $N=21$ ($D_A \approx
3.1$) while, on the contrary, even $10^7$ points are not yet enough in
the case $N=20$ ($D_A \approx 6.6$).  Indeed by inverting
(\ref{M-df}), we should expect $\epsilon_{\mathrm{min}} \propto
M^{-1/D_A}$, as shown in Fig.~\ref{epsM}.  The differences between the
case $N=21$ and $N=20$ in Fig.~\ref{corrX} and \ref{epsM} are thus a
mere consequence of the different attractor dimensionality, namely
$D_A(N=21)<D_A(N=20)$.

Relation (\ref{M-df}) also lays the basis for understanding the limits
of the Grassberger and Procaccia method\cite{Grassberger1983} to
compute the correlation dimension from the scaling behavior of the
correlation sum, or its approximation (\ref{grasspro}). In fact, it
states that the larger the dimension of the attractor the larger the
number of points $M$ necessary to sample it within a given accuracy
$\epsilon$.  For example, a segment of size $L$ will require $M \simeq
L/\epsilon$ points, $M \simeq (L/\epsilon)^2$ for a square and so
on. Smith\cite{Smith1988} proposed a minimum number of points of $M
\sim 42^{D_A}$ (about a decade and a half of scaling region) to get
reliable results.  For $D_A = 5$ or $6$, Smith's recipe requires from
hundreds of millions to billions of data, too large for standard
experiments. The above considerations on the limits of applicability
of the Grassberger and Procaccia\cite{Grassberger1983} technique may
sound trivial. However, in the '80s, when nonlinear time series
analysis started to be massively employed in experimental data
analysis, the limitations due the length of the time series were
overlooked by many researchers and a number of misleading papers
appeared even in important journals (for a critical review see
Ref.~\onlinecite{Ruelle1989}).

In conclusion, the possibility to predict the
future from the past using the analogues has its practical validity
only for low-dimensional systems.  More than one century after,
scientists working on prediction problems rediscovered 
Maxwell's warning: \textit{same antecedents never again concur,
and nothing ever happens twice}, whenever the system is moderately high
dimensional.

\subsection{Remarks on the case of unknown phase space\label{sec:embedding}}

So far, we have assumed that the vector ${\bm x}$ determining the
state of the system is known and can be measured with arbitrary
precision.  The real situation is less simple: usually, we do not know
the whole set of variables (not even their number) which define the
state of a system.  Moreover, even knowing them, in experimental
measurements, we normally have access only to very few scalar
observables $u_t$ depending on the state of the system: $u_t=G[{\bm
    x}_t]$.  In these cases, there exists a powerful technique (based
on Takens' \textit{delay embedding theorem}\cite{Takens1981}) able to
reconstruct the phase space, providing a rigorous ground to the use of
the analogues.\cite{Kostelich1994} Beyond the technical (often non
trivial) aspects, the main limit of the method, i.e. the exponential
increase of $M$ with $D_A$, still remains. Moreover, in practical
implementations, the presence of unavoidable measurement errors
introduces a further source of complications.  Ways to deal with the
general case of phase-space reconstruction and measurement errors have
been developed, however their discussion is beyond the scope of this
paper, so we refer the reader to specialized
monographs.\cite{Abarbanel,Kantz1997}

\section{Two examples where the method of analogues works\label{sec:5}}

Chaotic low dimensional attractors ($D_A \!\approx\!  2\!-\!4$) may
occur in many physical systems (such as electric circuits, lasers,
fluid motion etc., see Ref.~\onlinecite{Kantz1997}). 
Other natural phenomena, such as weather, are instead
characterized by high dimensional attractors with $D_A$ proportional
to the total number of variables involved, which is huge in the case
of the atmosphere. Thus, the conclusions of the previous section are
very pessimistic: when $D_A$ is that large only \textit{mediocre}
analogues (rather large $\epsilon$) can be found and those are, from
the point of view of predictability, usually not so informative about
the future evolution of the system.\cite{Lorenz1969a}

It is instructive, however, to consider here two exceptions to this
rule, namely: a variation on the theme of the Lorenz-96 model
(\ref{l96}) and briefly discuss tidal predictions which represent, to
the best of our knowledge, one of the few instances in which the idea
of using the past to predict the future works and has still important
practical applications.

\subsection{Systems with multiscale structure\label{sec:multi}}

We consider here systems with a multiscale structure, where the vector
state ${\bm x}$ can be decomposed into a slow component ${\bm X}$
which is also the ``largest'' one, and a fast component ${\bm y}$
``small'' with respect to ${\bm X}$ (i.e. $y_{\mathrm{rms}} \ll
X_{\mathrm{rms}}$). If the slow components can be described in terms
of an ``effective number'' of degrees of freedom much smaller than
those necessary to characterize the whole dynamics, mediocre (referred to
the whole system) analogues can be used to forecast at least the
slower evolving component.
As an illustration of such kind of system we consider a variant of the 
model (\ref{l96}) introduced by Lorenz himself\cite{Lorenz1996} to
discuss the predictability problem in the atmosphere, where indeed a
multiscale structure is present. The model reads  
\begin{eqnarray}
\strut{\hspace{-0.6truecm}}\frac{ \textrm{d} X_n} {\textrm{d} t} \!&=&\!  
X_{n-1}(X_{n+1}\!-\!X_{n-2})\!-\!X_n\!+\!F \!-\! \frac{hc}{b}
 \sum_{k=1}^{K} y_{k,n}    \label{lore2scaleA}\\
\strut{\hspace{-0.6truecm}}\frac{ \textrm{d} y_{k,n}} {\textrm{d} t} \!&=&\! 
cb\,  y_{k+1,n}(y_{k-1,n}\!-\!y_{k+2,n})\!-\!c\,y_{k,n} +\frac{hc}{b}  X_n
    \label{lore2scaleB}
  \end{eqnarray}
where $n=1,\ldots,N$ and $k=1,\ldots,K$ with boundary conditions $X_{N
  \pm n}=X_{\pm n}$, $y_{K+1,n}=y_{1,n+1}$ and
$y_{0,n}=y_{K,n-1}$. Equation~(\ref{lore2scaleA}) is essentially
(\ref{l96}) but for the last term which couples $\bm X$ to $\bm
y$. The variables $\bm y$ evolve with a similar dynamics but are
$c$ times faster and $b$ times smaller in amplitude. The parameter $h$,
set to $1$, controls the coupling strength.
\begin{figure}[hbt]
\includegraphics[clip=true,width=0.44\textwidth]{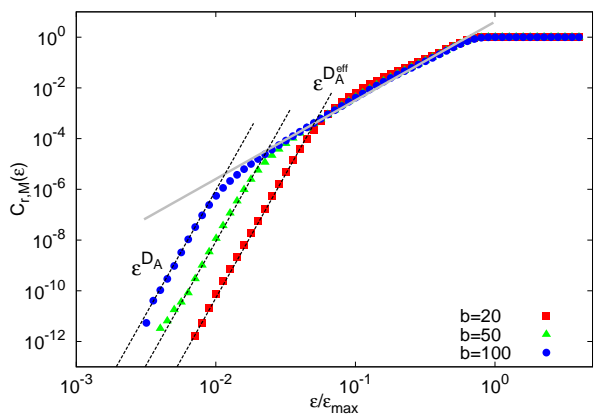}
\caption{(Color online) $C_{r,M}(\epsilon)$
  vs. $\epsilon/\epsilon_{max}$ for model
  (\ref{lore2scaleA}-\ref{lore2scaleB}) computed for three scale
  separations $b$ (as labeled) holding the other parameters fixed at
  $h=1$, $c=10$, $F=10$, $N=5$ and $K=10$. The gray straight line has
  slope $\approx 3.1$ while the dashed lines have all the same slope
  $\approx 9.8$.  The quantity (\ref{grasspro}) has been computed with
  $r=10^3$ and $M=10^7$.\label{multi} }
\end{figure} 

We repeat the computation to measure the probability of
$\epsilon$-analogues for the dynamics
(\ref{lore2scaleA}-\ref{lore2scaleB}), by assuming that the whole
state of the system $\bm x(t)=(\bm X(t),\bm y(t))$ is accessible, and
by ignoring which are the slow and fast variables, so that we must
search for the analogues in the sequence of states $\bm x_k=(\bm
X(t_k),\bm y(t_k))$, with $t_k=k\Delta t$.  Figure~\ref{multi} shows
$C_{r,M}(\epsilon)$ as a function of
$\epsilon/\epsilon_{\mathrm{max}}$ for a long sequence, $M=10^7$, for
fixed time scale separation $c=10$ and taking the fast component $\bm
y$ respectively $b=20, 50$ and $100$ times smaller than the slow one
$\bm X$. The phase-space dimensionality is $50$, with $N=5$ slow and
$K=10$ fast degrees of freedom. The attractor dimension of the whole
system $D_A$, given by the scaling $C(\epsilon) \sim \epsilon^{D_A}$
at very small $\epsilon$, is rather large ($D_A \approx 10$).
However, for $\epsilon/\epsilon_{\mathrm{max}} > O(1/b)$ we see a
second power law $C(\epsilon) \sim \epsilon^{D_A^{\mathrm{eff}}}$ with
$D_A^{\mathrm{eff}} \approx 3 < D_A$ which defines a sort of
``effective dimension at large scale''.

Therefore, if we are interested in predicting the slow evolving
component of the system, provided it is described by a relatively low
number of effective degrees of freedom, as here, we can exploit the
mediocre analogues (i.e. the $\epsilon$-analogues with
$\epsilon/\epsilon_{\mathrm{max}} > O(1/b)$). Moreover, with reference
to Eq.~(\ref{eq:lyap}), it is reasonable to expect that the prediction
error related to mediocre analogues grows as $ \sim \epsilon
e^{\lambda(\epsilon) T}$ where $\lambda(\epsilon)$ can be much smaller
than the Lyapunov exponent $\lambda_1$ (indeed as shown in
Ref.~\onlinecite{BOFFETTAETAL} $\lambda(\epsilon) \approx
\lambda_1/c$).  This implies that slow variables can be predicted over
longer term than the whole state of the system, as already realized by
Lorenz.\cite{Lorenz1996} In general multiscale systems, increasing
$\epsilon$ amounts to perform a coarse-graining on the system
description, which implies the ``elimination'' of the fastest degrees
of freedom, associated to the smallest scales. Consequently,
coarse-graining reduces the number of effective degrees of freedom
($D_A^{\mathrm{eff}}(\epsilon)<D_A$) and the error growth rate
($\lambda(\epsilon)<\lambda_1$).

The previous example is somehow the simplest multiscale system,
i.e. $C(\epsilon)$ vs $\epsilon$ shows only two logarithmic
slopes, $D_A^{\mathrm{eff}}$ and $D_A$. More generally one can have a
logarithmic slope $D(\epsilon)$ with a series of plateaux:
$D(\epsilon) \approx D^{\mathrm{eff}}_1$ for $\epsilon \in [\epsilon_0
  , \epsilon_1] $, $D(\epsilon) \approx D^{\mathrm{eff}}_{2} >
D^{\mathrm{eff}}_{1}$ for $\epsilon \in [\epsilon_1 , \epsilon_2] $,
and so on ($\epsilon_0 > \epsilon_1 > \epsilon_2 \dots $). The
interested reader may reproduce such a behavior by computing the
correlation integral of the discrete-time system discussed in
Ref.~\onlinecite{Olbrich}.

\subsection{Tidal prediction from past history}

Tidal prediction is a problem of obvious importance for navigation.
The appropriate governing equations had been established since long
time by Laplace: it is necessary to study the water level, with
suitable boundary conditions, under the gravitational forcing of the
Moon, the Sun and the Earth.\cite{Munk1966} Due to the practical
difficulties in the treatment of boundary conditions (only partially
known and hard to solve numerically), even with powerful computers the
fundamental equations cannot be directly used for tide forecasting.

However, and remarkably, already in the first half of the 19-th
century there existed efficient empirical methods to compile numerical
tables of tides in any location where a record of past tides were
known.\cite{Ducheyne2010} As recognized by Laplace, a great
simplification comes from the periodicity of the forcing (related to
the motions of celestial bodies) that naturally suggests to treat
tides in terms of Fourier series, whose frequencies are known from
celestial mechanics. Lord Kelvin and George Darwin (Charles' son)
showed that water levels can be well predicted by a limited number of
harmonics (say $10$ or $20$), determining the Fourier coefficients
from the past time data at the location of interest.  To make the
numerical computations automatic minimizing the possibility of error,
Kelvin and Darwin built a tide-predicting machine: a special-purpose
mechanical computer made of gears and pulleys. Tide-predicting
machines have been working till half century ago, when they were
replaced by digital computers to compute the Fourier
series.\cite{Ehret2008}

Since tides are chaotic,  it is natural to wonder why their
prediction from past records is a relatively easy task.
One realizes that the reason of such a favorable circumstance is
the low number of effective degrees of freedom involved.  In a
detailed description of tides also small scale phenomena are
involved, with very short characteristic times, \textit{e.g.}
micro-turbulence; therefore the ``true'' $D_A$ is surely very large,
together with $\lambda_1$. Therefore, the success of tidal prediction
is mainly a consequence of the multiscale character of the system,
that has a small $D^{\mathrm{eff}}$ (and also a small
$\lambda(\epsilon)$) on the interesting not too small scales, in a way
similar to the multiscale model of the previous subsection.  Indeed,
quite recently, investigations\cite{Khokhlov2008} of tidal time series
by using the standard method of nonlinear time series analysis (such
as embedding etc, see Sect.~\ref{sec:embedding}) found effective
attractor dimensions quite low (between $3$ and $4$) with effective
Lyapunov exponents of the order of $5$ days$^{-1}$. That explains a
posteriori the success of the empirical method.  Thanks to the low
$D^{\mathrm{eff}}$, analogues can be found.  Moreover, to forecast
tides a few hours in advance, the relatively low value of the Lyapunov
exponent makes the predictability time long enough for practical
purposes.  Of course, quantitative details (the precise values of
$D^{\mathrm{eff}}$ and of $\lambda (\epsilon)$) depend on the
locations,\cite{Frison1999} but for the method to work, the very
important aspect is the limited value of the effective attractor
dimension.

\section{Conclusions\label{sec:6}}

It is a common belief that chaos is the main limiting factor to
predictability in deterministic systems. This is correct as long as
the evolution laws of the system under consideration are
known. Conversely, if the information on the system evolution is
  only based on observational data, the bottleneck lies in Poincar\'e
recurrences which, in turn, depend on the number of effective degrees
of freedom involved.  Indeed, even in the most optimistic conditions,
if the state vector of the system would be known with arbitrary
precision, the amount of data necessary to make the predictions
meaningful would grow exponentially with the effective number of
degrees of freedom, independently of the presence of chaos. However,
when, as for tidal predictions, the number of degrees of freedom
associated with the scales of interest is relatively small,
future can be successfully predicted from past history.

We stress that, the necessity of an exponentially large (with $D_A$)
amount of data constitutes a genuine intrinsic difficulty of every
analysis based on time series without any guess on the underlying
dynamics.  Such a difficulty is not a peculiarity of the method of
analogues, but is inherent to all methods based on the occurrence
frequency of sequences of states to estimate the average of
observables.  In other words, the problem arises whenever one needs to
collect enough recurrences.  This obstacle may be partially overcome
by suitable information-theoretic techniques (see, e.g.,
Ref.~\onlinecite{Creutzig}) allowing for optimized reconstructions of
the dynamics, whose dimensionality, however, increases with the
required accuracy.  These conclusions are further supported by a
recent work by Cubitt and coworkers \cite{Cubitt}, showing that the
reconstruction of dynamical equations from data is a computationally
NP-hard problem, as the needed observation time scales exponentially
with the number of degrees of freedom.

In general, the best strategy for meaningful prediction is that
envisaged by Richardson, as a clever compromise between modeling and
data analysis. In this regard, we would like to conclude mentioning
that, in the era of information technology, the enormous capacity of
data storage, acquisition and elaboration may entitle someone to
believe that meaningful predictions can be extracted merely from data.
For example, recently the magazine Wired provocatively titled an
article ``The End of Theory: The Data Deluge Makes the Scientific
Method Obsolete'',\cite{wired} asserting that nowadays, with the
availability of massive data, the traditional way science progresses
by hypothesizing, modeling, and testing is becoming obsolete.  In this
respect, we believe that, while it is undeniable that the enormous
amount of data poses new challenges, the role of modeling cannot be
undermined.  When the number of effective degrees of freedom
underlying a dynamical process is even moderately large, predictions
based solely on observational data soon become problematic as it
happens in weather forecasting.

\section*{ACKNOWLEDGMENTS}
We thank U. Marini Bettolo Marconi for a careful reading of the
manuscript. We are grateful to an anonymous referee for useful remarks
and suggestions.

\end{document}